\documentclass[12pt,preprint]{aastex}

\slugcomment{July 6, 2006; accepted by the ApJ}

\shorttitle{Intermediate-mass Protostar MMS 7 in OMC-3}
\shortauthors{Takahashi et al.}

\begin{document}

%% LaTeX will automatically break titles if they run longer than
%% one line. However, you may use \\ to force a line break if
%% you desire.

\title{Millimeter- and Submillimeter-Wave Observations\\
of the OMC-2/3 Region;\\
I. Dispersing and Rotating Core around\\
an Intermediate-mass Protostar MMS 7\altaffilmark{1}}

%% Use \author, \affil, and the \and command to format
%% author and affiliation information.
%% Note that \email has replaced the old \authoremail command
%% from AASTeX v4.0. You can use \email to mark an email address
%% anywhere in the paper, not just in the front matter.
%% As in the title, use \\ to force line breaks.

\author{SATOKO TAKAHASHI}
\affil{Department of Astronomical Science, The Graduate University for Advanced Studies, 
National Astronomical Observatory of Japan, Osawa 2-21-1, Mitaka, Tokyo 181-8588, Japan; satoko.takahashi@nao.ac.jp}

\author{MASAO SAITO, SHIGEHISA TAKAKUWA and RYOHEI KAWABE}
\affil{ALMA Project Office, National Astronomical Observatory of Japan, Osawa 2-21-1, Mitaka, 
Tokyo 181-8588, Japan; Masao.Saito@nao.ac.jp, s.takakuwa@nao.ac.jp, ryo.kawabe@nao.ac.jp}

%% Notice that each of these authors has alternate affiliations, which
%% are identified by the \altaffilmark after each name.  Specify alternate
%% affiliation information with \altaffiltext, with one command per each
%% affiliation.

\altaffiltext{1}{Based on the observations made at the Nobeyama Radio Observatory, which is a branch of the National Astronomical Observatory,
an interuniversity research institute operated by the Ministry of Education, Culture, Sports, Science and Technology.}

%% Mark off your abstract in the ``abstract'' environment. In the manuscript
%% style, abstract will output a Received/Accepted line after the
%% title and affiliation information. No date will appear since the author
%% does not have this information. The dates will be filled in by the
%% editorial office after submission.

\begin{abstract}
We report the results of H$^{13}$CO$^{+}$(1--0), CO(1--0), and 3.3 mm dust continuum 
observations toward one of the strongest mm-wave sources in OMC-3, MMS 7, with the Nobeyama Millimeter Array (NMA) and the Nobeyama 45 m telescope.
With the NMA, we detected centrally-condensed 3.3 mm dust-continuum emission 
which coincides with the MIR source and the free-free jet.
%8 $\mu$m  and  24 $\mu$m SPITZER point sources and a 3.6 cm VLA source. 
The size and mass of the dusty condensation are 1500 $\times$ 1200 AU (P.A. $\sim$ 170$^{\circ}$) and 0.36 - 0.72 M$_{\rm{\odot}}$ 
(for $T_{\rm{dust}}$ = 26 - 50 K), respectively.
Our combined H$^{13}$CO$^{+}$ observations with the 45 m telescope and the NMA
have revealed a disk-like envelope around MMS 7 inside the H$^{13}$CO$^{+}$ core.
The size and the mass of the disk-like envelope are 0.15 $\times$ 0.11 pc and 
5.1 - 9.1 M$_{\rm{\odot}}$ (for $T_{\rm{ex}}$ = 26 - 50 K), respectively. 
The combined map also shows that the outer portion of the disk-like envelope has a fan-shaped structure 
which delineates the rim of the CO(1--0) outflow observed with the NMA.
%along the bipolar outflow direction. 
The position-velocity (P-V) diagrams in the H$^{13}$CO$^{+}$ (1--0) emission show that the velocity field 
in the disk-like envelope is composed of a dispersing gas motion and a possible rigid-like rotation. 
The mass dispersing rate is estimated to be
(3.4 - 6.0) $\times$ 10$^{-5}$ M$_{\rm{\odot}}$ yr$^{-1}$,
which implies that MMS 7 has an ability to disperse ${\sim}$10 M$_{\odot}$
during the protostellar evolutional time of a few $\times$ $10^{5}$ yr.
One of the probable dispersing mechanisms is the associated molecular outflow, 
and another the stellar wind which has enough power ($\sim$76 L$_{\odot}$) to drive the dissipation, (4.2 - 7.4) $\times$ 10$^{-3}$ L$_{\rm{\odot}}$.
The specific angular momentum of the possible rotation in the disk-like envelope
is nearly two orders of magnitude larger than that in low-mass cores. 
%at the same size scale.
The turn-over point of the power law of the angular momentum distribution in the disk-like envelope ($\leq$ 0.007 pc),
which is likely to be related to the outer radius of the central mass accretion,
is similar to the size of the 3.3 mm dust condensation.
We propose that the intermediate-mass protostar MMS 7 is in the last stage of the main accretion phase 
and that the substantial portion of the outer gas has already been dispersed, 
while the mass accretion may still be on-going at the innermost region traced by the dusty condensation.
%the outer dense gas has been already being dispersed, while at
%the innermost region traced by the dusty condensation mass accretion may be still ongoing.
\end{abstract}

%% Keywords should appear after the \end{abstract} command. The uncommented
%% example has been keyed in ApJ style. See the instructions to authors
%% for the journal to which you are submitting your paper to determine
%% what keyword punctuation is appropriate.


%% From the front matter, we move on to the body of the paper.
%% In the first two sections, notice the use of the natbib \citep
%% and \citet commands to identify citations.  The citations are
%% tied to the reference list via symbolic KEYs. The KEY corresponds
%% to the KEY in the \bibitem in the reference list below. We have
%% chosen the first three characters of the first author's name plus
%% the last two numeral of the year of publication as our KEY for
%% each reference.

%% Authors who wish to have the most important objects in their paper
%% linked in the electronic edition to a data center may do so by tagging
%% their objects with \objectname{} or \object{}.  Each macro takes the
%% object name as its required argument. The optional, square-bracket 
%% argument should be used in cases where the data center identification
%% differs from what is to be printed in the paper.  The text appearing 
%% in curly braces is what will appear in print in the published paper. 
%% If the object name is recognized by the data centers, it will be linked
%% in the electronic edition to the object data available at the data centers  

\section{Introduction}

In the last two decades, developments of millimeter and submillimeter interferometers 
have enabled us to establish 
%make high-sensitivity observations. In low-mass star-forming regions, 
%it has been revealed that there are gas accretions with $\dot{M}~{\sim}~10^{-5} - 10^{-6}~\rm{M_{\odot}~\rm{yr^{-1}}}$ 
%toward the central stars and circumstellar disks, 
%and active molecular outflows (e.g. Hayashi et al. 1993, Saito et al. 1996, Momose et al. 1998).
a standard scenario of low-mass star formation via disk accretion processes (e.g. Hayashi et al. 1993, Saito et al. 1996, Myers et al. 2000). 
%In addition, there are samples of outflowing shells with wide opening angles in low-mass cores,
%such as around IRAS 04368+2557 (L1527) \citep{oha97a}.
On the other hand, high- and intermediate-mass star-formation has not been clearly understood because of 
the complexity and the lack of observational studies, respectively. 
%Furthermore, the formation and evolution of intermediate-mass stars have not been well understood,  
%because there are limited observational studies. 
%Since intermediate-mass star formation is less 
%complicated, it would provide us a key to extend the low-mass scenario 
%and to bridge a missing link between the high-mass and low-mass star formation. 
%Interesting questions are if intermediate-mass protostars have higher mass-accretion or outflow 
%rates than those in low-mass cases. 
Since intermediate-mass star-formation is less complicated than high-mass star-formation, 
it would provide us a key to extend the low-mass scenario and to bridge a missing link between the high-mass and low-mass star formation. 
Particularly, it is important to verify the mechanisms and the differences of the 
destruction or dissipation of the natal dense cores between intermediate- and low-mass star formation,
because these phenomena are related to the termination of the mass accretion which determines
the final mass of the central star \citep{nak95}.
A key to these questions is investigation of structures and kinematics of dense cores around 
intermediate-mass protostars.
For these purposes, we have been conducting
survey observations toward protostellar cores in the Orion Molecular Cloud -2 and -3 region (OMC-2/3),
relatively close ($d$ = 450 pc; Genzel and Stutzki 1989) and 
the most representative intermediate-mass star-forming region.
We, here, report the results of one of the typical intermediate-mass protostars, 
MMS 7 in OMC-3, in the H$^{13}$CO$^{+}$(1--0), 
$^{12}$CO(1--0) and 3.3 mm continuum emission with the Nobeyama Millimeter Array (NMA) and the 
Nobeyama 45 m telescope. 
The H$^{13}$CO$^{+}$ (1--0) molecular line is one of the most appropriate tracers of dense gas,
which has a high critical density ($n_{\rm{crit}}~{\sim}~10^5~\rm{cm^{-3}}$) and 
is usually detected toward low- to intermediate-mass protostellar envelopes \citep{tak00, sai01, fue05}.

MMS 7 is 
%located at the northern part of an integral-shaped filament in the Orion A giant molecular cloud and is 
one of the Class 0 candidates identified by the 1.3 mm continuum observations toward OMC-3 \citep{chi97}. 
This object is also identified as CSO12 by the 350 $\mu$m continuum observations \citep{lis98} 
and is associated with the IRAS point source 05329-0508.
%H$^{13}$CO$^{+}$(1-0) observations were also carried out by Aso et al. (2000) toward the OMC-2/3 region with the Nobeyama 
%45 m telescope and substantial dense gas in MMS 7 was detected. 
The dust mass derived from the 1.3 mm continuum emission and the bolometric luminosity are 8 M$_{\odot}$ and 76 L$_{\odot}$, respectively \citep{chi97}. 
This bolometric luminosity corresponds to $\sim$ 3 M$_{\odot}$ or a A0 star at the ZAMS.
Dense molecular gas around MMS 7 is also detected by H$^{13}$CO$^{+}$(1--0) observations \citep{aso00}.
MMS 7 is also associated with a bright reflection nebula, Haro-5a/6a \citep{har53}, 
and 2MASS and mid-infrared sources are located at the root of the eastern reflection nebula \citep{nie03}. 
At MMS 7, a giant molecular outflow (0.86 pc; Aso et al. 2000) is also observed along the east-west direction, 
and 2.12 $\mu$m H$_2 ~v$=1-0 $S$(1) knots are detected up to 1.45 pc to the western side \citep{yu97, sta02}. 
In addition, there is a 3.6 cm continuum source elongated along the large-scale outflow, which traces a free-free jet from the protostar. 
The relative isolation of MMS 7 along with the above properties makes this source one of the most appropriate objects to investigate detailed 
spatial and velocity structures of a dense core around an intermediate-mass protostar.

%The H$^{13}$CO$^{+}$ (1--0) molecular line is one of the most appropriate tracers of dense gas,
%which has a high critical density ($n_{\rm{crit}}~{\sim}~10^5~\rm{cm^{-3}}$) and 
%is usually detected toward low- to intermediate-mass protostellar envelopes \citep{oni00, tak00, fue05}.
%H$^{13}$CO$^{+}$(1-0) observations of dense gas in the Taurus Molecular Cloud complex have been carried out by Saito et al. (2001). 
%They revealed that the dense gas dissipates as protostars evolve and classified the evolutional phase from class A to C.
%H$^{13}$CO$^{+}$(1-0) observations were also carried out by Aso et al. (2000) toward the OMC-2/3 region with the Nobeyama 
%45 m telescope and substantial dense gas in MMS 7 was detected. 
%Our observations with the higher spatial resolution and higher sensitivity enable us to discuss details of the structure and 
%kinematics of the dense gas around the intermediate-mass protostar of MMS 7 on 10$^3$-10$^4$ AU scale .
 
% \clearpage

\section{Observations and Data Reduction}

%% In a manner similar to \objectname authors can provide links to dataset
%% hosted at participating data centers via the \dataset{} command.  The
%% second curly bracket argument is printed in the text while the first
%% parentheses argument serves as the valid data set identifier.  Large
%% lists of data set are best provided in a table (see Table 3 for an example).
%% Valid data set identifiers should be obtained from the data center that
%% is currently hosting the data.

\subsection{NMA Observations}

We have observed H$^{13}$CO$^{+}$ ($J$=1--0; 86.754 GHz) and CO ($J$=1--0; 115.271 GHz) lines 
in MMS 7 with the six-elements NMA  from 2004 November to 2005 March. 
Detailed NMA observational parameters are summarized in Table \ref{obs}.
Since the minimum projected baseline lengths of the H$^{13}$CO$^{+}$(1--0) and CO(1--0) observations were 2.9 and 6.3 k$\lambda$, 
our observations were insensitive to structures more extended than 57$''$ (0.13 pc) and 32$''$ (0.07 pc) at the 10 \% level, respectively \citep{wil94}.
%The field center was set to be the peak position of the 1.3 mm dust continuum emission 
%[R.A. = 05 h 35 m 26.4 s, Dec = -05 d 3 m 53.4 s (J2000)]. 
%The primary beam sizes of the 10m dishes (FWHM of the field of view) are 77 $''$ at 87 GHz and 62 $''$ at 115 GHz.  
%The H$^{13}$CO$^{+}$ and CO images were obtained with the D + C and D configuration, respectively.  
% The ranges of the projected baseline length were 2.9 k$\lambda$ -115 k$\lambda$ at 87 GHz and 
%6.3 k$\lambda$ -27.4 k$\lambda$ at 115 GHz, and
%our observations were insensitive to structures extending more than 57$''$ at 86 GHz and 32$''$ at 115 GHz at the 10 \% level \citep{wil94}.
%The DSB system noise temperatures of the SIS receiver were 150 - 300 K at 87GHz and 100 - 400 K at 115 GHz toward the zenith. 
%The backend was a 1024-channel digital spectrocorrelator, 
%FX, with a total bandwidth of 32 MHz. 
%The velocity resolution of the H$^{13}$CO$^{+}$ (1--0) and $^{12}$CO (1--0) lines was
%0.11 km s$^{-1}$ and 0.08 km s$^{-1}$, respectively.
%The response across the observed pass-band for each sideband 
%was determined from 40 minutes observations of 3C454.3, 3C273 and 0423-013. 
%A gain calibrator, 0530+135, was observed every 20 minutes. 
%The flux scale of 0530+135 was derived from the flux of the secondary calibrator, 
%3C454.3 and 3C84, whose fluxes were determined by the observations of the primary calibrator (i.e. Uranus). 
 The overall flux uncertainty was estimated to be $\sim$ 15 percent. 
 After the calibrations, we made final images only from the data taken under good weather conditions. 
 We have CLEANed images with a natural weighting by using Astronomical Image Processing System (AIPS) 
 developed at NRAO
 \footnote{NRAO is facility of the National Science Foundation operated under cooperative 
 agreement by Associated Universities, Inc.}.
%The synthesized beam has a size of 6$''$.77 $\times$ 4$''$.34 (3000$\times$2000 AU) with 
 %a position angle of -21.35$^{\circ}$ in the H$^{13}$CO$^{+}$ maps and 6$''$.04$\times$4$''$.74 (2700$\times$2100 AU) 
 %with a position angle of -11.42 $^{\circ}$ in the CO maps. 

 The 3.3 mm continuum data with the D, C, and AB configurations was obtained simultaneously with the H$^{13}$CO$^{+}$ data 
 by using a digital spectral correlator UWBC \citep{oku00}, which has 128 frequency channels and a 1024 MHz bandwidth per baseline. 
 Both the lower (87 $\pm$ 0.512 GHz) and upper (99 $\pm$ 0.512 GHz) sidebands were obtained simultaneously 
 with a phase-switching technique. 
 To obtain a higher signal-to-noise ratio, the data of both sidebands were added (effective frequency = 92 GHz).  
 We have CLEANed the continuum data with a uniform weighting (see Table \ref{CONT_T} and Section 3.1).

 \subsection{45 m Observations and Combining with the NMA Data}

Single-dish mapping observations of MMS 7 in the H$^{13}$CO$^{+}$ (1--0) line 
were also conducted with the 5$\times$5 SIS array receiver BEARS equipped
in the 45-m telescope on 2005 April. 
At 87 GHz, the half-power beamwidth (HPBW) and the main-beam efficiency of the 45 m antenna were 18$''$.5 and 0.5, respectively. 
We observed the $190 '' \times 190 ''$ area (corresponding to 0.43 pc$\times$0.43 pc) 
centered on the NMA field center with a grid spacing of 8$''$.2. 
The temperature scale was determined by a chopper-wheel method, 
which provides us the antenna temperature corrected for the atmospheric attenuation.
As a backend, we used an auto correlator with a frequency resolution of 31.12 kHz, corresponding to 0.11 km s$^{-1}$ at the H$^{13}$CO$^{+}$ frequency. 
We applied scaling factors to correct for variations of the gain and the DSB ratio in each beam, 
which were measured by the Nobeyama Radio Observatory and the typical value was $\sim$1.5.
The typical system noise temperature in DSB mode was 240 K at the elevation of 70$^{\circ}$. 
The typical noise level was 0.1 K in $T_{\rm{A}}^{\ast}$. 
The telescope pointing was checked every $\sim$ 90 minutes by five-point cross scans 
in the SiO($v$=1,$J$=1--0) maser emission 
from Orion KL [RA. = 5 h  35 m 14.5 s  Dec = -05 d 22 m 30.4 s (J2000)]
using the 40-GHz SIS receiver. 
The pointing error ranged from 1$''$ to 5 $''$ during the observing run.

% \subsection{Combining the NMA and  the 45 m Data in the H$^{13}$CO$^{+}$(1--0) line}
After the calibrations of the 45 m and NMA data, we combined each data set in the 
H$^{13}$CO$^{+}$(1--0) line 
using MIRIAD (we followed the instruction presented by Takakuwa et al. 2003a)
to fill the missing short-spacing information of the interferometric data 
from the single-dish data and to obtain more feasible images. 
%The antenna temperature of the 45 m data was converted to fluxes with a scaling factor of 1 K = 4.0 Jy/18.5$''$ beam, 
%which takes the main beam-efficiency into account. 
Our data taken with the 45 m telescope and the NMA show a discrepancy of the amplitude scale at the overlapping $uv$ distance (2.8 - 6.2 k$\lambda$).
In order to compensate this discrepancy, we multiply the 45 m amplitude by a scaling factor of 1.4, 
which was derived from the amplitude of these $uv$ data at the overlapping $uv$ distance. 
We consider that causes of the discrepancy are the degradation of the main-beam efficiency under windy conditions 
and the uncertainty of the DSB ratio of BEARS. 
We have CLEANed the combined images by uniform weighting. 
The combined synthesized beam has a size of $5''.81 \times 3''.95$ (2600$\times$1800 AU) with a position angle of -24.26 $^{\circ}$. 
Figure \ref{comb} compares the integrated intensity maps created with the 45m data (middle), NMA data (bottom left)
and the combined map (bottom right). 
By combining the single-dish and the interferometric data, we succeeded to remove deep negative bowls seen in the NMA map 
(bottom left) and recover extended emission in the dense-gas filament,  
which is important to discuss the spatial and velocity structure without any problem of missing fluxes.
 
%% In this section, we use  the \subsection command to set off
%% a subsection.  \footnote is used to insert a footnote to the text.

%% Observe the use of the LaTeX \label
%% command after the \subsection to give a symbolic KEY to the
%% subsection for cross-referencing in a \ref command.
%% You can use LaTeX's \ref and \label commands to keep track of
%% cross-references to sections, equations, tables, and figures.
%% That way, if you change the order of any elements, LaTeX will
%% automatically renumber them.

%% This section also includes several of the displayed math environments
%% mentioned in the Author Guide.

\section{RESULTS}
\subsection{Millimeter Continuum Images of the Inner Dusty Condensation}

Figure \ref{cont} shows the 3.3 mm continuum emission observed with the NMA
(white contours) superposed on the 2MASS K's-band image.
Different NMA maps in Figure \ref{cont} were made 
from different sets of the visibility data whose parameters are listed in Table \ref{CONT_T}. 
We have detected the intense 3.3 mm continuum emission toward MMS 7.
%which is considered to mainly come from thermal radiations of dust grains around the protostar. 
Although the peak position of the 3.3 mm source appears to be shifted from the 1.3 mm emission peak  by 2.7$''$ (Chini et al. 1997; See Figure 2), 
the positions of these two sources are consistent within the expected positional error of the 1.3 mm observations (${\pm}5''$) \citep{chi97}.
%The peak position of the 3.3 mm continuum source is roughly consistens with the 1.3 mm emission peak (i.e. field center) 
%observed by the IRAM-30 m telescope, $\sigma{\sim}$2$''$.7 $\pm$ 0$''$.3, (Chini et al. 1997; See Figure 2).
%The positions of these two sources are consistent within the expected position error of 1.3 mm observations ($\pm 5''$) from Chini et al. 1997.
The 3.3 mm source is located at the root of the K's reflection nebula and 
also coincides with the 8 and 24 $\mu$m compact source observed 
with {\it SPITZER/IRAC} and the 3.6 cm ionized jet \citep{rei99,rei04} 
and therefore, this source is considered to be a deeply embedded protostar. 
 
On the assumption that the dust emission is optically thin at 3.3 mm and the distribution 
of the dust temperature is uniform,
we can estimate the mass of the dusty condensation by 

\begin{equation}
M_{\rm{dust}} = \frac{F_{\rm{\lambda}} d^2}{ \kappa_{\rm{\lambda}} B_{\lambda} (T_{\rm{dust}})},
\end{equation}

where $\kappa_{\lambda}$ is a mass-absorption coefficient of dust grains, 
$B_{\lambda}(T)$ is a Planck function, $F_{\lambda}$ is a total flux density of the continuum 
emission and $d$ is a distance to MMS 7 (450 pc; Genzel and Stutzki 1989). 
The total flux of the 3.3 mm continuum emission, 23.82 mJy, includes the contribution from the free-free jet.
On the assumption that the spectral index of the free-free jet is ${\sim}0.6$ \citep{ang98, rey86}, 
the expected flux density of the free-free emission at 3.3 mm becomes 2.47 mJy based on the 3.6 cm flux density of 0.59 mJy \citep{rei99}. 
Thus, the flux density attributed to the dust emission is 21.35 mJy at 3.3 mm. 
%When the spectral index of free-free jet, 0.6, is assumed (Anglada et al. 1998, Reynolds 1986), 
%the effect of free-free emission at 3.3 mm becomes 2.47 mJy at 3.3 mm based on the 3.6 cm flux density, 0.59 mJy.
%Then F$_{\lambda}$ = 21.35 mJy.
Adopting the dust opacity of $\kappa_{\lambda} = 0.037~\rm{cm^2 g^{-1}}~(\lambda/400 \rm{\mu m})^{\beta}$  
\citep{oha96}, $\beta$ = 1.0 which is calculated from the 1.3 mm and 3.3 mm total fluxes and $T_{\rm{dust}}$ = 26 - 50 K \citep{chi97, joh03}, 
we estimate the mass of the dusty condensation to be 0.36 - 0.72 M$_{\odot}$. 

Figure \ref{cont}$c$ was made only from the visibility data taken at $>$ 30 k$\lambda$ (See Table 2). 
In the higher resolution image, we see an elongated structure along the NE-SW direction.
This NE-SW elongation may imply the presence of a binary with 
a separation of $\sim$ 1$''$.9 (corresponds to 830 AU).
However, there is no infrared or cm counterpart toward the extension of the 3.3 mm emission, and
more plausible interpretation is that the faint continuum feature traces the hot and 
dense region heated by the associated outflow,
since the direction of the elongation of the 3.3 mm emission is similar to the direction of the 
3.6 cm fine-scale jet.

\subsection{Bipolar Outflow associated with MMS 7} 

Figure \ref{INT}$a$ shows the distribution of the blueshifted ($V_{\rm{LSR}}$=7.0 to 8.7 km s$^{-1}$) 
and redshifted ($V_{\rm{LSR}}$=12.1 to 13.8 km s$^{-1}$) $^{12}$CO(1--0) 
emission in MMS 7 superposed on the 2MASS K's-band image. 
Here, we adopt the systemic velocity of 10.6 km s$^{-1}$ obtained from the single-dish H$^{13}$CO$^{+}$(1--0) spectrum. 
The outflow is elongated along the east-west direction and roughly aligned with the direction of the free-free jet \citep{rei99}.
Both the blue- and red-shifted components are seen at the east and west side of MMS 7, 
which suggests that the axis of the bipolar outflow is near the plane of the sky.
Furthermore, the outflow shows an asymmetric structure with the stronger red- 
and blue-shifted components at the eastern side.

Figure \ref{12CO_ch} shows velocity channel maps in the $^{12}$CO(1--0) emission with a velocity interval of 0.6 km s$^{-1}$. 
The red-shifted component in the velocity range of $V_{\rm{LSR}}$=11.6 to 13.3 $\rm{km~s^{-1}}$ was already detected with the BIMA and the FCRAO 14 m telescope \citep{wil03}. 
The distribution of this component is consistent with that found by their observations.
We have newly detected other components in the velocity range from $V_{\rm{LSR}}$=7.6 to 11.0 $\rm{km~s^{-1}}$. 
High-velocity blue-shifted components in the velocity range of $V_{\rm{LSR}}$=7.6 to 8.7 $\rm{km~s^{-1}}$ are seen at the eastern side of MMS 7. 
Particularly, the strong and compact CO component is evident at the position of the K's-band nebula which is most likely illuminated by MMS 7.
The strong blueshifted and redshifted components are only seen at the eastern side.
These results suggest that the $^{12}$CO(1--0) bipolar outflow is slightly inclined from the plane of the sky 
and the western components are hidden by the disk-like envelope, 
i.e. the eastern blue-shifted components is located near-side.
This is supported by the fan-shaped brighter reflection nebula at the eastern side than that at the western side (See Figure 1). 
Figure \ref{12CO_ASTE} shows the distribution of the high-velocity $^{12}$CO(3--2) emission observed with 
the ASTE telescope (Atacama Submillimeter Telescope Experiment). 
The large-scale outflow ($\sim $1 pc) also has the same configuration as that of the small-scale outflow observed with the NMA in the $^{12}$CO(1--0) emission, 
i.e. the blue lobe is located at the eastern side and the red lobe at the western side.
The detailed interpretation of the CO(3--2) outflow taken with the ASTE telescope will be the subject to the forthcoming paper.

\subsection{H$^{13}$CO$^{+}$($J$=1--0) emission }

Both the Single-dish (Figure \ref{comb}$a$) and interferometric (Figure \ref{comb}$b$) maps in the H$^{13}$CO$^{+}$(1--0) emission 
show an fan-shaped structure, which is approximately perpendicular to the associated bipolar outflow, 
and is also seen in the single-dish 1.3 mm and 350 $\mu$m continuum emission \citep{chi97, lis98}.
%total intensity maps taken with the 45 m has a cavity-like structure 
%(dashed line in Figure \ref{comb}$a$) whose orientation roughly coincides with the direction of 
%the $^{12}$CO bipolar outflow.
The interferometric image, Figure 1$b$, shows an elongated disk-like envelope associated with MMS 7 
(we use the term ``disk-like envelope'' as the compact structure detected with the NMA).
The size of the disk-like envelope is estimated to be 30000$\times$21000 AU (0.15$\times$0.11 pc).
%The disk-like envelope shows a fan-shaped structure 
The combined image, Figure 1$c$, reveals that the disk-like envelope observed with the NMA 
is embedded in the extended dense gas halo observed with the 45 m telescope.

Figure 3$b$ and $c$ compare the morphology of the disk-like envelope to that of the 2 $\mu$m and 8 $\mu$m emission, respectively.
The distribution of the extended near- and mid-infrared emission seems to be anti-correlated with that of the disk-like envelope. 
Particularly, the local emission maximum in the disk-like envelope correspond to the local minimum in the 8 $\mu$m map and vice versa.
The central point source in the 8 $\mu$m emission, which coincides with the 3.3 mm point source and therefore the protostellar source of MMS 7,  
is not located at the peak of the disk-like envelope  and is shifted to the eastern side (see Figure {\ref{INT}$c$). 
One interpretation to explain this offset is that the warm and dense dusty condensation in the vicinity of the protostar does not reside at the 
peak of the molecular column density traced by the H$^{13}$CO$^{+}$ emission.

Table \ref{H13_T} summarizes physical properties of dense gas around MMS 7 derived from the 45m and NMA combined data.
Assuming that the H$^{13}$CO$^{+}$ emission is optically thin and 
the distribution of the excitation temperature is uniform, we estimated the LTE mass as follows,
\begin{equation}
\textstyle{
	M_{\rm{LTE}}~[\rm{M}_{\odot}] = 2.03{\times}10^{-2} \left( \frac{26~\rm{K}}{T_{\rm{ex}}} \right) 
	\exp \left(\frac{4.16}{T_{\rm{ex}}}\right) \frac{\tau}{1-\exp({-\tau})} \left(\frac{d}{450~{\rm{pc}}}\right)^2
	\left( \frac{1.4{\times}10^{-10}}{X \left({\rm H}^{13}{\rm{CO}}^{+}\right)} \right) \\
	\int F_{\nu} dv~[\rm{Jy~km~s}^{-1}],
}
\end{equation}
 We adopt $T_{\rm{ex}}$ =26-50 K \citep{chi97, joh03},
 and $X$ [H$^{13}$CO$^{+}$] $\sim$ 1.4$\times$10$^{-10}$ derived from the comparison of the estimated 
 H$^{13}$CO$^{+}$ column density and the C$^{18}$O column density \citep{chi97} .

Figure \ref{H13_ch} shows NMA velocity channel maps in the  H$^{13}$CO$^{+}$ (1-0) emission at a velocity interval of 0.26 $\rm{km~s^{-1}}$. 
The blue-shifted emission, $V_{\rm{LSR}}$=9.3-9.8 km s$^{-1}$,
is seen at the south-western part of the 3 mm dust peak. 
Close to the systemic velocity of $V_{\rm{LSR}}$=10.6 km s$^{-1}$,
the peak position of the blue-shifted emission shifts from south-west to west. 
The red-shifted emission, $V_{\rm{LSR}}$=10.8-11.5 km s$^{-1}$, is seen at the northern part of the 3 mm dust peak.
This velocity gradient along the major axis of the disk-like envelope implies the rotational motion.
We also note the velocity gradient along the minor axis, 
i.e. the strong blue-shifted component in the velocity range of $V_{\rm{LSR}}$=10.2-10.4 km s$^{-1}$ is located at the western part of the protostar, 
while there appears weak red-shifted extension in the velocity range of $V_{\rm{LSR}}$=10.6-11.3 km s$^{-1}$ at the eastern side of the protostar.
The fan-shaped structure in the total integrated intensity map of Figure{\ref{comb}}$b$
is also confirmed in the velocity channel maps 
in the range of $V_{\rm{LSR}}$ =  10.2 -10.4 and 10.8-11.3 km s$^{-1}$.

\subsection{Another YSO candidate; MMS 7-NE}

 We have also detected a weak 3.3 mm continuum source at 3.8 $\sigma$ level toward $\sim$ 40$''$ northeast 
 of MMS 7 (see Figure \ref{cont}$a$; hereafter we call this component MMS 7-NE). 
 The peak position and the total 3.3 mm flux of this source are estimated to be 
 RA= 05$^h$35$^m$28.2$^s$, Dec= -05$^{\circ}$03$\arcmin$41.1$\arcsec$ (J2000) and 6.81 mJy, 
 respectively, from the 2-dimensional Gaussian fitting to the image.
 This source is also seen as a point source in the K's image,
 and 8$\mu$m and 24$\mu$m sources taken with the {\it SPITZER/IRAC} (see Figure 3).
This source was also detected to be a very compact 350 $\mu$m dust component
by Lis et al. (1988) with CSO,
which locates at the southeast of CSO11,
although they did not identify this object.
 The 3.3 mm continuum flux corresponds to 0.04-0.30 M$_{\odot}$ 
 \footnote{No cm source was detected toward this source in the cm observations by Reipurth et al. (1999).
Thus, we estimated the upper limit of the possible contribution from the free-free emission at 3.3 mm.
Given the rms noise level of 40 ${\mu}$Jy, the 3 sigma upper limit of the 3.6 cm total flux (3 sigma level)
is 0.12 mJy. Thus, if we assume the spectral index of 0.6, the contribution from the free-free emission at 3.3 mm is
0.49 mJy, which is less than 5\% of the total flux of the 3.3 mm continuum emission at MMS 7-NE (6.81 mJy).
%Considering the absolute flux uncertainty of 10\% in the 3.3 mm observations, 
We can probably ignore the possible contribution. 
%There is no cm source in this region. 
%When the rm noise level, 40 micro Jy, is adopted the upper limit of the 3.6 cm total flux (Reipurthe et al. 1998), 
%the contribution from free-free emission at 3.3 mm is 0.16 mJy $<<$ 2.7 mJy (total flux of 3mm continuum emission at MMS 7-NE ).
%So, we can neglect the free-free emission.
}
 assuming $\beta=0-1,~T_{\rm{dust}}~=~20~\rm{K}$ and gas-to-dust ratio of 100. 
 
MMS 7-NE is associated with the weak H$^{13}$CO$^{+}$ emission (Figure 1$b,c$ or Figure 3).
The $^{12}$CO emission is also seen toward MMS 7-NE (see Figure \ref{12CO_ch} in the range 
of $V_{\rm{LSR}}$ = 8.2 \rm{to} 10.4 km s$^{-1}$),
implying the association of the molecular outflow.
We note that MMS 7-NE has a smaller scale (10$^{4}$ AU scale) outflow,
suggesting that it has little influence to destruct the MMS 7 disk-like envelope.

MMS 7-NE is likely to be a more evolved protostar than MMS 7,
because there is a diffraction ring in the 8 $\mu$m image which suggests the presence of point source 
(i.e. no scattering material) and a K's-band counterpart of this source. 
Alternatively, the circumstellar material around MMS 7-NE has already been swept up by the outflow from MMS 7, 
revealing the central stellar object of MMS 7-NE.
%% information to the copy editor.  This information will appear as a
%% footnote on the printed copy for the manuscript style file.  Nothing will
%% appear on the printed copy if the preprint or
%% preprint2 style files are used.

%% The eqnarray environment produces multi-line display math. The end of
%% each line is marked with a \\. Lines will be numbered unless the \\
%% is preceded by a \nonumber command.
%% Alignment points are marked by ampersands (&). There should be two
%% ampersands (&) per line.

%% Putting eqnarrays or equations inside the mathletters environment groups
%% the enclosed equations by letter. For instance, the eqnarray below, instead
%% of being numbered, say, (4) and (5), would be numbered (4a) and (4b).
%% LaTeX the paper and look at the output to see the results.

%% This section contains more display math examples, including unnumbered
%% equations (displaymath environment). The last paragraph includes some
%% examples of in-line math featuring a couple of the AASTeX symbol macros.

\section{DISCUSSION}

 Our new observations in the H$^{13}$CO$^{+}$ line have revealed systematic velocity structures 
 in the disk-like envelope around the intermediate-mass protostar of MMS 7.
 In the subsequent sections, we will discuss the gas kinematics in the disk-like envelope and the difference 
 from that in low-mass counterparts using position-velocity (P-V) diagrams.

\subsection{Radial Velocity Structure of the Disk-like Envelope}
 \subsubsection{P-V diagram}
 
We show a P-V diagram along the minor axis of the disk-like envelope in Figure \ref{H13_PV}$b$. 
In the P-V diagram, there are at least two components; 
one is an eastern component at the velocity range of $V_{\rm{LSR}}$=10.7 to 11.5 km s$^{-1}$ ((i) in Figure \ref{H13_PV}$b$) 
and the other western-component at the velocity of $V_{\rm{LSR}}$=9.6 to 10.9 km s$^{-1}$.
The western component is associated with the 3 mm dust continuum emission (vertical solid line in Figure 7b). 
In this component, ``X-shape'' velocity structures are discerned in the P-V diagram.
One of the X-shaped velocity structures as indicated by (ii) in Figure \ref{H13_PV}$b$ shows blue-shifted emission ($V_{\rm{LSR}}$=9.6 to 10.0 km s$^{-1}$) 
at the west of the protostar and red-shifted emission ($V_{\rm{LSR}}$=10.6 to 10.9 km s$^{-1}$) at the east of the protostar.
The sense of this velocity structure is opposite to that of the associated molecular outflow.
On the assumption that the H$^{13}$CO$^{+}$ emission traces the flattened disk-like envelope which is perpendicular to the associated outflow, 
this velocity structure can be interpreted as an expanding motion in the flattened disk (e.g. Kitamura et al. 1996).
The other velocity structure in the X-shape, with two components at the velocity range of $V_{\rm{LSR}}$=9.9 to 10.4 km s$^{-1}$ 
(blueshifted; (iii)-1 in Figure \ref{H13_PV}$b$) and $V_{\rm{LSR}}$=10.5 to 11.0 km s$^{-1}$ (redshifted; (iii)-2 in Figure \ref{H13_PV}$b$), 
has the same velocity sense as that of the associated outflow. These two components are located outside the 
central condensed gas, suggesting that these components are not associated with the central protostar.
Thus, it is natural to interpret that these gas components trace swept-up dense gas by the associated outflow perpendicular to the disk-like envelope.
A similar example has also been reported from observations in L1228 \citep{taf94}.
They detected a velocity shift in the high-density tracer of the C$_3$H$_2$ (2$_{12}$ - 1$_{01}$) emission, 
which has the same velocity sense as that of the associated CO bipolar outflow, and they interpreted this shift
as an interaction between the dense core and the high-velocity outflow.
The eastern component, indicated by (i) in Figure 7, is located at 25$''$ east of the continuum peak position
and is a part of the fan-shaped structure seen in the combined total intensity map. 
We interpret that this component is remnant dense gas interacting with the associated molecular outflow. 

In summary, the P-V diagram can be interpreted as a mixture of the dispersing gas along the disk-like envelope and the interacting dense gas 
with the associated outflow perpendicularly to the disk-like envelope. Presumably due to the insufficient spatial resolution, 
an infall gas motion in the protostellar source of MMS 7 has not been clearly identified in our observations.
 
\subsubsection{Properties of the Dispersing Disk}
 
The velocity structure of the disk-like envelope shows a dispersing gas motion as discussed in the last section.
The virial mass of the disk-like envelope is estimated to be 23-30 M$_{\odot}$
assuming $D_{\rm{env}}$ = 0.15 pc, $T_{\rm{env}}$ = 26 - 50 K, 
and $\Delta v$ = 1.0 km s$^{-1}$, which is larger than the LTE mass of the disk-like envelope (5.1 - 9.1 M$_{\odot}$).
This result suggests that the disk-like envelope is gravitationally unbound, 
supporting our interpretation of the dispersing gas motion.
We can estimate physical parameters of the dispersing motion, such as an expanding velocity 
($V_{\rm{exp}}$), momentum ($P_{\rm{exp}}$), 
expanding energy ($E_{\rm{exp}}$), and mechanical power ($L_{\rm{exp}}$), as  $V_{\rm{exp}}({\equiv}~V_{\rm{max}})~=~1.2~\rm{km~s^{-1}}$, 
$P_{\rm{exp}}=MV_{\rm{max}}~\rm{M_{\odot} km~s^{-1}}$ ,
$E_{\rm{exp}}=MV_{\rm{max}}^2/2~\rm{M_{\odot} km ^{2}~s^{-2}}$, 
and $L_{\rm{exp}}=MV_{\rm{max}}^3/2R~\rm{L_{\odot}}$.
We adopt the disk inclination angle from the plane of the sky to be $i~\sim$ 80$^{\circ}$,
which is consistent with the morphology of the CO outflow and the reflection nebula.
Table \ref{COMP_T} lists the estimated physical parameters of the dispersing motion.
The dispersing envelope has also been observed in the $^{13}$CO(1-0) emission around
a low-mass YSO of DG Tau \citep{kit96}.
We compare physical parameters specified in the dispersing process of MMS 7 to those of DG Tau in Table \ref{COMP_T}.
These parameters of MMS 7 are two orders of magnitude larger than those in DG Tau. 
The interferometric observations of DG tau may suffer from the problem of the missing flux, 
which prevents us from making a direct comparison of these parameters.
However, even if the interferometric observations recover only 10 \% of the total flux from DG Tau, 
these parameters of MMS 7 are still one order of magnitude larger than those in DG Tau.
These results suggest that the intermediate-mass protostar of MMS 7 
has more active dispersing processes than low-mass counterparts.
 
There are recent studies on the dispersing gas motion around Herbig Ae/Be stars \citep{fue02}. 
Their studies show that intermediate-mass stars disperse $\gtrsim$ 90 \% of the total mass of the parent core during the protostellar phase. 
Particularly, these objects which have a spectral type later than B6 evolve with the substantial dispersing material, 
3 - 80 M$_{\odot}$, in the protostellar phase. 
We, here, define the mass-dispersing rate (mass loss rate) which is 
$\dot{M}_{\rm{out}}\sim M_{\rm{env}}{\cdot}V_{\rm{exp}}/R_{\rm{env}}=~(3.4 -6.0){\times}10^{-5}~ \rm{M_{\odot}~yr^{-1}}$ . 
This value implies that MMS 7 has an ability to disperse a mass of 0.9-1.5 M$_{\odot}$ 
during the outflow dynamical time scale of 2.5$\times$10$^4$ yr,
which is estimated from the ASTE data.
We speculate that the intermediate-mass protostar of MMS 7 will keep dispersing
substantial circumstellar material during the protostellar phase
and will eventually be a Herbig Ae star in $\tau \leq$ a few$\times$10$^{5}$ yr, 
as proposed by Fuente et al. (2002).

\subsubsection{Driving Mechanisms of the Dispersing Disk}

What is the mechanism of the dispersing process in the disk-like envelope ?
One of the probable candidates is the associated bipolar outflow, which is blowing away the material along the outflow axis and 
at the surface of the disk-like envelope.
There is a twisting structure near the center of the reflection nebula Haro-5a/6a in the K's-band seen in Figure \ref{INT}, 
and several knots (at least three components) in the red-shifted CO(3-2) emission in Figure \ref{12CO_ASTE} (a),(b) denoted by the dashed-blue line.
The connected trajectory of those emission peaks shows a wiggled structure.
These results suggest that the precessing bipolar outflow from MMS 7 is able to destruct 
the surface of the disk-like envelope in the east-west direction with a wide opening angle. 
The importance of bipolar outflows in the dispersal of ambient gas around low-mass protostars has already been 
pointed out by previous studies \citep{oha97a, vel98, tak03b}.
Takakuwa et al. (2003b) has reported a dispersing low-mass protostellar core around IRAM 04191+1522. 
The detected blueshifted CH$_{3}$OH components
(see BLUE1 and BLUE2 of Figure 3 in Takakuwa et al. 2003b) 
are most likely to be formed in consequence of the interaction between the outflow 
and the ambient dense gas surrounding the protostar and pushed away from the natal cloud core.

In addition, there is a possibility that the stellar wind from the central protostar mainly contributes to the dispersion along the direction 
which is perpendicular to the outflow.
A low-velocity wind ($\sim$100 km s$^{-1}$) with a wide opening angle, ($\sim$100$^{\circ}$) is detected toward a low-mass protostar L1551 IRS 5  \citep{pyo02, pyo05}.
There are also studies on the stellar wind around Herbig Ae/Be stars, which show the derived mass-loss rate in  the ionized gas 
$\sim$10$^{-8}$ - 10$^{-7}$ M$_{\odot}$ yr$^{-1}$ (e.g. Skinner et al. 1994).
Therefore, it is possible that the stellar wind from MMS 7 drives the mass dispersion with the estimated rate of (3.4 - 6.0)$\times$10$^{-5}$ M$_{\odot}$ yr$^{-1}$.
In fact, the bolometric power of MMS 7 is 76 $\rm{L_{\odot}}$, much larger than the mechanical luminosity of the dispersing motion, 
$L_{\rm{exp}}=(4.2 - 7.4){\times}10^{-3}~\rm{L_{\odot}}$ (see Table \ref{COMP_T}), and is energetically sufficient to drive the expanding motion.

\subsection{Rotating Envelope}

The P-V diagram along the major axis in Figure \ref{H13_PV}$c$ shows that the velocity of the disk-like envelope 
increases with distance up to 5200 AU from the center. 
%One of the possible interpretations is the global motion of the OMC filament.
%Because of systemic velocity of MMS 7, 10.6 km s$^{-1}$, shifted to the other OMC-3 sources in the filament, $\sim$11.3 km s$^{-1}$.
One of the possible (and simple) interpretations of this velocity gradient is rigid rotation in the disk-like envelope, 
although, it is difficult to clearly discern it owing to the mixture of the kinematics of the filament/core/envelope.
%We, now, assume this velocity gradient as a rigid-rotation in following discussion.
%This velocity structure is likely to suggest the presence of the rigid rotation in the disk-like envelope.
Assuming the disk inclination of $80^{\circ}$ and the radius of $5200$ AU, 
the timescale of the rigid rotation in this flattened disk is estimated to be $1.5 \times 10^5$ yr, 
which is much longer than the outflow dynamical time scale of 2.5$\times$10$^4$ yr.
 
Red and blue plots in Figure \ref{ang} show the local specific angular momentum 
in the disk-like envelope as a function of the radius around MMS 7. 
Our data demonstrate  the angular momentum distribution on a size scale of $0.007 -0.02$ pc 
(corresponding to 1500 - 4700 AU) inside the single source. 
For comparison, we also plot the angular momentum of low-mass 
dense cores and circumstellar disks (Ohashi et al. 1997b, Goodman et al. 1993).
The value of the specific angular momentum around the MMS 7 is nearly two orders of magnitude larger than 
that of the low-mass counterparts at the same size scale.
The specific angular momentum distribution of the disk-like envelope around MMS 7 
has a power-law index of $\sim$ 1.8 which is closer to the index of rigid rotation, that is, 2.0.
On the other hand, the specific angular momenta of low-mass NH$_{3}$ cores observed by Goodman et al. (1993) 
are fitted by a power-law index of $\sim$ 1.6 as indicated by a dashed line in Figure {\ref{ang}}.
Furthermore, we note that there is no clear turn-over point of the power-law in the plot of MMS 7 
as the low-mass case,
which suggests that the turn-over point is smaller than that of low-mass NH$_{3}$ cores 
($\sim$ 0.03 pc)
and is smaller than 0.007 pc, on the assumption that the value of the specific angular momentum at the
innermost part of MMS 7 is the same as that of the low-mass value.
The turn over point of the power law indicates the minimum size of the rotation 
where the specific angular momentum is not constant as a function of radius. 
At the dynamical collapsing region, the specific angular momentum should be constant in the entire radius.
Therefore, the turn-over point can be considered as a starting point of the dynamical collapse toward the central protostar 
(Ohashi et al. 1997b).
From these considerations, we suggest that the infalling radius in MMS 7 is smaller than that of low-mass counterparts.
If dynamical collapse in MMS 7, if any, follows the inside-out collapse model \citep{shu77},  
the smaller infalling radius implies that the period passed after the accretion started is 
shorter than that of low-mass cores (a few $\times$10$^{5}$ yr).

%The other possibility to explain this velocity structure is the global velocity gradient in the cloud flament.
%Because of systemic velocity of MMS 7, 10.6 km s$^{-1}$, shifted to the other OMC-3 sources in the filament, $\sim$11.3 km s$^{-1}$.
%In the subsesquent papers, we will also discuss the global velocity structure of OMC-2/3 filament with the extensive 45 m observations.

\subsection{Evolutional Stage of the Intermediate-mass Protostar MMS 7}

The disk-like envelope, with the fan-shaped structure and the expanding motion around MMS 7, 
has a larger dispersing activity  than that of low-mass cores (see section 4.1). 
MMS 7 has a large-scale (${\sim}$1 pc) active CO outflow (Figure \ref{12CO_ASTE})
which has a dynamical time scale of $\tau_{\rm{dyn}}{\sim}2.5{\times}10^{4}$ yr.
These observational results suggest the presence of the significant mass-loss activity
in the intermediate-mass protostar of MMS 7.
Then, is there any mass supply onto the central star from the surrounding gas around MMS 7 ?
In this subsection, we will discuss a possibility of the presence of mass accretion and the evolutionary stage of MMS 7.

Although an infall motion is not clearly seen in our H$^{13}$CO$^{+}$ observations,
the presence of the molecular outflow implies the existence of the mass accretion
toward MMS 7 \citep{bon96}.
To make a 3 M$_{\odot}$ protostar of MMS 7, 
a large accretion rate  ($\dot{M} \sim 1.2 \times 10^{-4}~\rm{M_{\odot}~yr^{-1}}$) is 
required over the outflow dynamical time 
(we, here, assume that $\tau_{\rm{dyn}}{\sim} \tau_{\ast}$ and constant $\dot{M}$). 
We found the smaller turn-over radius of the specific angular momentum distribution, 
which implies that the radius of the infalling region, $\leq$ 0.007 pc (1500 AU),  is smaller than low-mass counterparts.
The size of the dusty condensation around MMS 7,  1500 $\times$ 1200 AU,  is similar to the inferred radius of the infalling region.
Thus, it is possible that inside the dusty condensation, 
where the present spatial resolution is not high enough to trace the internal velocity structure, 
there is a substantial material which is accreting onto the central protostar.
These considerations suggest that the intermediate-mass protostar MMS 7 
is formed with higher mass-accretion and mass-loss rates than those of low-mass counterparts.

The small infalling radius if present, short dynamical time of the outflow, and the presence of substantial circumstellar material, 
could imply that MMS 7 is in the early evolutional phase.
On the other hand, our observational results show the energetic dispersing activity.
Therefore, the early phase intermediate-mass protostar MMS 7 is already dispersing the surrounding
envelope.
Future interferometric observations with higher spatial resolution is required to 
investigate details of the innermost accreting region around MMS 7.

\section{
}

We have carried out H$^{13}$CO$^{+}$ (1--0), CO (1--0) and 3.3 mm dust-continuum observations 
toward the intermediate-mass protostar of MMS 7 in OMC-3 with the NMA and the NRO 45 m telescope. 
Figure \ref{schem} shows the schematic picture of MMS 7 region and the main results are summarized as follows:

\begin{enumerate}
	\item	We detected a centrally-condensed 3.3 mm continuum source which is consistent with the 1.3 mm and SPITZER source.
				The size and mass of the centrally-condensed component are 1500 $\times$ 1200 AU (P.A. $\sim$ 170$^{\circ}$) and 0.36 - 0.72 M$_{\odot}$, respectively.
				This 3.3 mm source is most likely to be an associated dusty condensation around the intermediate-mass protostar of MMS 7.
				We have also identified a new 3.3 mm continuum source, MMS 7-NE, whose mass is estimated to be 0.04 - 0.30 M$_{\odot}$.
%				toward $\sim 40''$ northeast of MMS 7. The mass of the dust
%				condensation is estimated to be 0.04 - 3.2 M$_{\odot}$ ($\beta$= 0 - 2).
%				MMS 7-NE is likely to be a more evolved protostar than MMS 7.
				
	\item	The H$^{13}$CO$^{+}$ emission is distributed 
				surrounding the 3.3 mm dust-continuum emission and has fan-shaped structure at the rim of the the bipolar outflow, suggesting the presence of the interacting region.
				Our combined 45 m and NMA data have revealed a disk-like envelope inside the dense H$^{13}$CO$^{+}$ core.
				The size and the mass of the disk-like envelope are 0.5 $\times$ 0.11 pc (P.A. $\sim$ 0$^{\circ}$) and 
				5.1 - 9.1 M$_{\rm{\odot}}$ (for $T_{\rm{ex}}$ = 26 - 50 K), respectively.   
				
	\item	We found that the disk-like envelope around MMS 7 is being dispersed.
				The momentum, expanding energy, and mechanical power of the dispersing envelope are much larger than those of low-mass dispersing envelopes.
				The disk-like envelope has a quite large mass-dispersing rate of (3.4 - 6.0) $\times$ 10$^{-5}~ \rm{M_{\odot}~yr^{-1}}$ 
				and this result implies that MMS 7 has an ability to disperse the substantial mass ($\sim$ 10 M$_{\odot}$) 
				during the protostellar evolutional phase ($\sim$ a few$\times$10$^{5}$ yr).
				The mechanism of the dispersing gas motion is considered to be the effect of the associated CO bipolar outflow 
				and the stellar wind from the protostar. 
				
	\item	We also detected a velocity gradient in the disk-like envelope from an P-V diagram along the major axis. 
				The specific angular momentum inside the disk-like envelope is 
				nearly two orders of magnitude larger than that in low-mass NH$_{3}$ cores at same size scale.
				The power-law index of the angular momentum distribution inside the envelope is estimated to be 1.8, 
				which is closer to the index of a rigid rotation (= 2.0).
				In addition, we note that a turn-over point of the power-law in the disk-like envelope is $\leq$ 0.007 pc.
				This result suggests that the infalling radius is smaller than 0.007 pc, which is similar to the size of the 3.3 mm dust condensation.
				The presence of the significant mass-loss activity (dispersing envelope and molecular outflow) implies the presence of the mass supply 	
				onto the central star from the surrounding gas around MMS 7, i.e.  MMS 7 is formed with 
				the higher mass-accretion and mass-loss rate that those of low-mass counterparts.
 \end{enumerate}

%% The displaymath environment will produce the same sort of equation as
%% the equation environment, except that the equation will not be numbered
%% by LaTeX.

%% If you wish to include an acknowledgments section in your paper,
%% separate it off from the body of the text using the \acknowledgments
%% command.

%% Included in this acknowledgments section are examples of the
%% AASTeX hypertext markup commands. Use \url without the optional [HREF]
%% argument when you want to print the url directly in the text. Otherwise,
%% use either \url or \anchor, with the HREF as the first argument and the
%% text to be printed in the second.

\acknowledgments

The authors are grateful to the NRO staff for their support during the observations. 
We acknowledge R. Kandori and the SIRIUS team for providing us the NIR image taken by SIRIUS, 
D. Iono and L. Allen for helping us to archive SPITZER data, and N.Kobayashi for fruitful comments. 
We also thank to the referee for the constructive comments that have helped to improve this muniscript.
S. Takahashi is financially supported by the Japan Society for the Promotion of Science (JSPS) for Young Scientists.

%% To help institutions obtain information on the effectiveness of their
%% telescopes, the AAS Journals has created a group of keywords for telescope
%% facilities. A common set of keywords will make these types of searches
%% significantly easier and more accurate. In addition, they will also be
%% useful in linking papers together which utilize the same telescopes
%% within the framework of the National Virtual Observatory.
%% See the AASTeX Web site at http://www.journals.uchicago.edu/AAS/AASTeX
%% for information on obtaining the facility keywords.

%% After the acknowledgments section, use the following syntax and the
%% \facility{} macro to list the keywords of facilities used in the research
%% for the paper.  Each keyword will be checked against the master list during
%% copy editing.  Individual instruments or configurations can be provided 
%% in parentheses, after the keyword, but they will not be verified.

%% Appendix material should be preceded with a single \appendix command.
%% There should be a \section command for each appendix. Mark appendix
%% subsections with the same markup you use in the main body of the paper.

%% Each Appendix (indicated with \section) will be lettered A, B, C, etc.
%% The equation counter will reset when it encounters the \appendix
%% command and will number appendix equations (A1), (A2), etc.

\appendix

%% The reference list follows the main body and any appendices.
%% Use LaTeX's thebibliography environment to mark up your reference list.
%% Note \begin{thebibliography} is followed by an empty set of
%% curly braces.  If you forget this, LaTeX will generate the error
%% "Perhaps a missing \item?".
%%
%% thebibliography produces citations in the text using \bibitem-\cite
%% cross-referencing. Each reference is preceded by a
%% \bibitem command that defines in curly braces the KEY that corresponds
%% to the KEY in the \cite commands (see the first section above).
%% Make sure that you provide a unique KEY for every \bibitem or else the
%% paper will not LaTeX. The square brackets should contain
%% the citation text that LaTeX will insert in
%% place of the \cite commands.

%% We have used macros to produce journal name abbreviations.
%% AASTeX provides a number of these for the more frequently-cited journals.
%% See the Author Guide for a list of them.

%% Note that the style of the \bibitem labels (in []) is slightly
%% different from previous examples.  The natbib system solves a host
%% of citation expression problems, but it is necessary to clearly
%% delimit the year from the author name used in the citation.
%% See the natbib documentation for more details and options.

\clearpage

%% Use the figure environment and \plotone or \plottwo to include
%% figures and captions in your electronic submission.
%% To embed the sample graphics in
%% the file, uncomment the \plotone, \plottwo, and
%% \includegraphics commands
%%
%% If you need a layout that cannot be achieved with \plotone or
%% \plottwo, you can invoke the graphicx package directly with the
%% \includegraphics command or use \plotfiddle. For more information,
%% please see the tutorial on "Using Electronic Art with AASTeX" in the
%% documentation section at the AASTeX Web site,
%% http://www.journals.uchicago.edu/AAS/AASTeX.
%%
%% The examples below also include sample markup for submission of
%% supplemental electronic materials. As always, be sure to check
%% the instructions to authors for the journal you are submitting to
%% for specific submissions guidelines as they vary from
%% journal to journal.
%% This example uses \plotone to include an EPS file scaled to
%% 80% of its natural size with \epsscale. Its caption
%% has been written to indicate that additional figure parts will be
%% available in the electronic journal.

\clearpage

\clearpage

\begin{table}
\begin{center}
\caption{Parameters for the NMA Observations \label{obs}}
\begin{tabular}{lcc}
\tableline\tableline
\multicolumn{1}{l}{}  & \multicolumn{2}{c}{Molecular Line}  \\
 \cline{2-3}
\multicolumn{1}{l}{Parameter}  & {H$^{13}$CO$^{+}$(1--0)}  & {CO(1--0)}   \\
 \tableline
Configuration \tablenotemark{a}	&	D, C and AB	&	D	\\
Baseline [k$\lambda$]				 		& 2.9 -- 115 &	6.3 -- 27.4	 \\
Phase reference center (J2000)				&	\multicolumn{2}{c}{R.A. = 05$^{h}$ 35$^{m}$ 26.4$^{s}$, Dec = -05$^{\circ}$ 03$'$ 53.4$''$}    \\
Primary beam HPWB [arcsec]			&	77$''$			&	62$''$		\\
Synthesized Beam HPBW [arcsec]	&	6$''$.77 $\times$ 4$''$.33 (P.A.=-21$^{\circ}$.35) &	6$''$.04 $\times$ 4$''$.74 (P.A.=-11$^{\circ}$.42) \\
Velocity resotion [km s$^{-1}$]			&	0.11 km s$^{-1}$ 	&	0.08 km s$^{-1}$ \\
Band width [MHz] 							&	\multicolumn{2}{c}{32}	  \\
Gain calibrator	\tablenotemark{b}									&	\multicolumn{2}{c}{0531+135}	  \\
Passband calibrator	\tablenotemark{c}							&	3C454.3, 0423-013	&	3C273	  \\
Flux calibrator	\tablenotemark{d}									&	3C454.3, 0531+134	&	3C454.3	\\
System temperature in DSB [K]	\tablenotemark{e}	&		150 - 300 	&	100 - 400 \\
Rms noise level [Jy beam$^{-1}$] 	&		0.69		&		0.48	\\
 \tableline
\end{tabular}
\tablenotetext{a}{D and AB are the most compact and sparse configurations, respectively.}
\tablenotetext{b}{A gain calibrator, 0530+135, was observed every 20 minutes.}
\tablenotetext{c}{The passband observations were archived by 40 minute observations of each calibrator.}
\tablenotetext{d}{The flux of each calibrations was determined by the observations of the primary calibrator, Uranus.}
\tablenotetext{e}{The system noise temperature of SIS receiver were measured toward the zenith.}
\end{center}
\end{table}

\begin{table}
\begin{center}
\caption{Parameters of the Figure 2(a),(b) and 2(c)\label{CONT_T}}
\begin{tabular}{lrr}
\tableline\tableline
Parameter & Figure 2(a),(b) & Figure 2(c) \\
\tableline
Baseline [k$\lambda$]			&2.9 - 115  & 30 - 115 \\
Weighting 								& Uniform & Uniform \\
Beamsize (HPBW) [arcsec]	&$2''.47 \times 2''.04$ & $1''.83 \times 1''.60$  \\
P.A. of the beam [$^{\circ}$]			&-10.86 &-1.79	\\
Rms noise level [Jy beam$^{-1}$]	 & 8.37e-04 &1.23e-03 \\
Total flux density [mJy] ($> 3 \sigma$)	&	23.82 & 21.20 \\
Deconbolution size [arcsec]	& $3.40(\pm 0.185) \times 2.68(\pm 0.146)$	& $2.86(\pm 0.299) \times 2.29(\pm 0.239)$ \\
P.A. of the component	[$^{\circ}$] & $170.621 \pm 9.2$	& $161.262 \pm 18.464$ 	\\
\tableline
\end{tabular}
%% Any table notes must follow the \end{tabular} command.
%\tablenotetext{a}{Sample footnote for table~\ref{tbl-2} that was
%generated with the \LaTeX\ table environment}
%\tablenotetext{b}{Yet another sample footnote for table~\ref{tbl-2}}
%\tablenotetext{c}{Another sample footnote for table~\ref{tbl-2}}
%\tablecomments{We can also attach a long-ish paragraph of explanatory
%material to a table.}
\end{center}
\end{table}

%% If the table is more than one page long, the width of the table can vary
%% from page to page when the default \tablewidth is used, as below.  The
%% individual table widths for each page will be written to the log file; a
%% maximum tablewidth for the table can be computed from these values.
%% The \tablewidth argument can then be reset and the file reprocessed, so
%% that the table is of uniform width throughout. Try getting the widths
%% from the log file and changing the \tablewidth parameter to see how
%% adjusting this value affects table formatting.

%% The \dataset{} macro has also been applied to a few of the objects to
%% show how many observations can be tagged in a table.

\begin{table}
\begin{center}
\caption{Physical Properties of the H$^{13}$CO$^{+}$ (1-0) Components\label{H13_T}}
\begin{tabular}{lcccc}
\tableline\tableline
                & Size \tablenotemark{c} & $N_{\rm{H_2}}$\tablenotemark{d} & $M_{\rm{LTE}}$ \\
Component & (AU)  & ($\times 10^{22}$ cm$^{-2}$) & (M$_{\odot}$) \\
\tableline
Disk-like envelope\tablenotemark{a} & 30000 $\times$ 21000 & 7.3 (13.0) & 5.1 (9.1)\tablenotemark{d}  \\
MMS 7-NE\tablenotemark{b} &2600 $\times$ 1800 &2.4 (4.3)  & 6.6$\times10^{-2}$ ($1.2{\times}10^{-1})$\tablenotemark{e}  \\
\tableline
\end{tabular}
\tablenotetext{a}{We use a flux above 9 $\sigma$ level in the 45 m and NMA combined map }
\tablenotetext{b}{We assume that the H$^{13}$CO$^{+}$ gas extends similar to beamsize. }
\tablenotetext{c}{Assuming that $d$=450 pc}
\tablenotetext{d}{Derived from $T_{\rm{ex}}$ =26 K (50 K) and  $X[\rm{H^{13}CO^{+}}]=1.4{\times}10^{-10}$(Chini et al. 1997, Johnstone et al. 2003)} 
\tablenotetext{e}{Derived from $T_{\rm{ex}}$ =20 K and  $X[\rm{H^{13}CO^{+}}]=1.4{\times}10^{-10}$} 
\end{center}
\end{table}

\begin{table}
\begin{center}
\caption{Physical Parameters of the Dispersing Core \label{COMP_T}}
\begin{tabular}{lccc}
\tableline\tableline
\multicolumn{1}{l}{Parameter}  & \multicolumn{2}{c}{MMS 7\tablenotemark{a}} & \multicolumn{1}{l}{DG Tau\tablenotemark{b}} \\
 \tableline
$M_{\rm{env}} (\rm{M_{\odot}})$	&	5.1\tablenotemark{c}	&	9.1\tablenotemark{d}	&	0.03	\\
$P_{\rm{exp}}$ (M$_{\odot}$ km s$^{-1}$) & 6.2  &	11.1	& 0.05 \\
 $E_{\rm{exp}}$ (M$_{\odot}$ km$^{2}$ s$^{-2}$)  &	3.8	&	6.8  & 0.03   \\
 $L_{\rm{exp}}$ (10$^{-3}~\rm{L_{\odot}}$ )  & 	4.1	&	7.4	 & 0.6 \\
 \tableline
\end{tabular}
\tablenotetext{a}{Assuming that disk inclination is 80$^{\circ}$}
\tablenotetext{b}{Reffer from Kitamura et al. 1996 observed with the NMA data}
\tablenotetext{c}{Derived from $T_{\rm{ex}}$ = 26 K (Chini et al. 1997)}
\tablenotetext{d}{Derived from $T_{\rm{ex}}$ = 50 K (Johnstone et al. 2003)}
\end{center}
\end{table}

\clearpage

%% Text for table notes should follow after the \enddata but before
%% the \end{deluxetable}. Make sure there is at least one \tablenotemark
%% in the table for each \tablenotetext.
%\tablecomments{Table \ref{tbl-1} is published in its entirety in the 
%electronic edition of the {\it Astrophysical Journal}.  A portion is 
%shown here for guidance regarding its form and content.}
%\tablenotetext{a}{Sample footnote for table~\ref{tbl-1} that was generated
%with the deluxetable environment}
%\tablenotetext{b}{Another sample footnote for table~\ref{tbl-1}}
%\tablerefs{
%\end{deluxetable}

\end{document}